# A threshold-free summary index for quantifying the capacity of covariates to yield efficient treatment rules

**Authors:** Mohsen Sadatsafavi[1,2], Mohammad Mansournia[3], Paul Gustafson[4]
1. Faculty of Pharmaceutical Sciences, University of British Columbia, Vancouver, British Columbia, Canada
2. Faculty of Medicine, University of British Columbia, Vancouver, British Columbia, Canada
3. Department of Epidemiology and Biostatistics, School of Public Health, Tehran University of Medical Sciences, Tehran, Iran
4. Department of Statistics, University of British Columbia, Vancouver, British Columbia, Canada

**Corresponding author:**   Mohsen Sadatsafavi
Faculty of Medicine,
The University of British Columbia
7th Floor, 828 West 10th Avenue
2405 Wesbrook Mall
Vancouver, BC Canada V6T 1Z3
Tel: 604.875.5178 | Fax: 604.875.5179
Email: msafavi@mail.ubc.ca

Table count: 2
Figure count: 2
Running head: A summary index for subgroup analysis in clinical trials
Word count: 4,588 (without abstract, appendix, references, figures, and tables)
Number of appendices: 1

**Source of support:** This study was supported by the Canadian Institutes of Health Research (Grant #155554).

**Conflict of interest:** None declared

## Abstract

The focus of this paper is on quantifying the capacity of covariates in devising efficient treatment rules when data from a randomized trial are available. Conventional one-variable-at-a-time subgroup analysis based on statistical hypothesis testing of covariate-by-treatment interaction is ill-suited for this purpose. The application of decision theory results in treatment rules that compare the expected benefit of treatment given the patient's covariates against a treatment threshold. However, determining treatment threshold is often context-specific, and any given threshold might seem arbitrary at the reporting stages of a clinical trial. We propose a threshold-free metric that quantifies the capacity of a set of covariates towards finding individuals who will benefit the most from treatment. The construct of the proposed metric is comparing the expected outcomes with and without knowledge of covariates when one of a two randomly selected patients are to be treated. We show that the resulting index can also be expressed in terms of integrated treatment benefit as a function of covariates over the entire range of treatment thresholds. We also propose a semi-parametric estimation method suitable for out-of-sample validation and adjustment for optimism. We use data from a clinical trial of preventive antibiotic therapy for reducing exacerbation rate in Chronic Obstructive Pulmonary Disease to demonstrate the calculations in a step-by-step fashion. The proposed index has intuitive and theoretically sound interpretation and can be estimated with relative ease for a wide class of regression models. Beyond the conceptual developments presented in this work, various aspects of estimation and inference for such metrics need to be pursued in future research.

**Keywords:** Subgroup Analysis; Prediction Modeling; Clinical Trials; Regression Analysis

**Introduction**

Clinical trials are major undertakings to evaluate the merits of medical interventions and are gateways to their market entry. Besides estimating average treatment effect, evaluating heterogeneity of treatment effect across identifiable subgroups is a common practice in the reporting stage of trials: 61% of all trials published during one year in a major medical journal reported on at least one such subgroup analysis(1). In addition to providing insight into the underlying disease processes, subgroup analysis can provide evidence as to whether the treatment should be provided to a subset of patients who will benefit the most from it(2).

The classical approach towards subgroup analysis in clinical trials is through statistical significance testing for the presence of covariate-by-treatment interaction(3). The problems with this approach are well recognized and extensively discussed(4). In brief, one issue is that type I and II error rates are not strictly under the control of the investigator for exploratory subgroup analysis. Another is that statistical testing of one covariate at a time makes it difficult to compare the benefit of treatment between two individuals as they likely differ in several aspects. Moreover, neither the level of statistical significance, nor the magnitude of the coefficient for the interaction term, the latter being dependent on the scale of the covariate, can be used to compare covariates in terms of their relative contribution towards treatment effect heterogeneity. Further, the presence of interaction on the 'scale of estimation' for subgroup analysis (often a relative scale) does not mean its presence on the 'scale of interest' for decision-making (often the absolute scale), or vice versa(2).

In light of these issues, there has been a call towards a 'risk-based' approach for subgroup analysis(4). In such an approach, covariates are combined into a single (externally or internally developed) risk score(4). The predicted risk score is then treated as the subgroup-defining variable. More recently, direct estimation of treatment benefit, the difference in the outcome with and without treatment as a function of the covariates of interest, is proposed(2,5). VanderWeele et al explored how covariates can be used to formulate optimal treatment rules when data on treatment assignment, covariates, and treatment outcomes are available(5). They examined a variety of objective functions and showed that they all result in applying a threshold on the expected difference in outcomes with and without the treatment given the patient's observed characteristics.

While such a move towards theoretically sound subgroup analysis is appealing, the practicality of a full decision-theoretic approach in clinical trials can be questioned. Maximizing commonly defined decision-theoretic objective functions requires weighting the short- and long-term outcomes of alternative treatment decisions. This requires a deep contextual investigation which can be difficult especially for new interventions. Reporting on the weighting mechanism might be seen as a distraction from the main findings of the clinical trial.

We are motivated by the approach taken by the marker development community facing a similar problem. Often, biomarkers and clinical prediction models need to be coupled with a positivity rule to enable binary classifications, such as labelling an individual as diseased versus healthy. Classical findings from decision theory identify the optimal positivity rule based on the consequences of test results(6,7). However, it is argued that in the early stages of marker development, applying such decision rules to find the optimal threshold can be challenging or even off-putting(8). Instead, the interest is focused on threshold-free, 'global', measures of discriminatory capacity of the marker, such as the Area Under the Curve (AUC) of the Receiver Operating Characteristic curve and the closely related Concordance (C) statistic, or the Integrated Discrimination Improvement index(9). The implicit promise is that if a marker performs well (or better than another marker) on these metrics, it is likely to have some value (or more value compared to another marker) when a specific threshold is applied.

In our opinion, subgroup analysis for clinical trials should similarly remain detached from the potentially contentious weighting of the outcomes. While a full decision-theoretic approach can take place in its due course, trialists can focus on 'AUC-like' metrics that quantify the overall capacity of covariates towards concentrating treatment benefit. In this work we propose such an index. Our proposed index is relatively easy to calculate, has intuitive interpretation, and does not suffer from many of the issues that hamper interpretations based on statistical significance of interaction terms. Our focus in this paper is on the conceptual foundations, and we relegate important issues regarding the merits of model selection, variable selection, estimation, and validation methods to future developments.

## Notation and context

We focus on randomized trials comparing two interventions among $n$ individuals. By $A$ we define the treatment variable, with $A = 1$ indicating treatment and $A = 0$ indicating no treatment. By $\mathbf{X}$ we refer to the set of covariates of interest for subgroup analysis. We define $Y_0$ and $Y_1$ as the (possibly counterfactual) outcomes for each patient under no treatment and treatment, respectively. The observed outcome, denoted by $Y$, is $(1 - A).Y_0 + A.Y_1$. We assume that observation on one unit is unaffected by the particular assignment of treatments to the other units(10). Our objective is to devise a metric that quantifies how $\mathbf{X}$ can help us find individuals with the highest expected treatment benefit on the decision scale, which we define to be $b = E(Y_1 - Y_0)$ if the outcome is favorable (e.g., 5-year survival) or $b = E(Y_0 - Y_1)$ if the outcome is unfavorable (e.g., disease recurrence). Without loss of generality, in what follows we assume that the outcome is an unfavorable event and treatment labels are such that the population-average treatment benefit is positive.

Our derivations start by estimating the expected benefit of treatment as a function of covariates. In most cases, this function is based on a parametric 'risk model' that estimates $E(Y|\mathbf{X}, A)$, the rate or risk of the outcome as a function of covariates and treatment. The expected treatment benefit for the i[th] subject can then be estimated as

$$\hat{b}_i = \hat{E}(Y_i|\mathbf{X}_i, A_i = 0) - \hat{E}(Y_i|\mathbf{X}_i, A_i = 1).$$

The coefficients of such a regression model can be fitted through maximum likelihood (ML) estimation. However, our objective is not to examine the causal relation between covariates and outcomes; rather, the interest is to investigate the performance of covariate(s) in predicting treatment outcomes in future patients. In this context, severe biases can arise when the same dataset is used for both ML parameter estimation and evaluation of the predictive performance of the model(11). It is known that regularization techniques, such as shrinking the parameter estimates towards the null, can improve the predictive performance in a new sample(12,13). The use of ensemble methods (such as the super-learner(14)) can also be considered, but the nuances of feature selection and estimation is not the primary focus of this work.

**To what extent covariates can help formulate an efficient treatment rule?**

If a universal treatment decision (treating all or treating none) will be applied to the population, the knowledge of patient covariates will be irrelevant. Such knowledge matters only if the treatment decision is moved from the population level to the individual level. To quantify the value of such knowledge, we contrast the outcomes of population-level versus individual-level treatment decisions through an abstract comparison that has a broad analogy with the concept behind the C statistic.

Imagine the task is front of us is to give treatment to one, and only to one, subject among a randomly selected pair of subjects. We compare the efficiency of this task with and without knowing about covariates. When we have knowledge of covariates, the most efficient treatment rule is the the one that provides treatment to the subject with higher expected treatment benefit (or at random if the estimates are tied) given the value of covariates. The average benefit of such a covariate-informed rule (compared with not treating any of the two subjects) will therefore be $E\{max(B_1, B_2)\}$, where $B_1$ and $B_2$ are random draws from the distribution of $b$. On the other hand, without knowledge of covariates, no rule is any more (or less) efficient than random treatment assignment between the two subjects. The average benefit of such a covariate-agnostic treatment rule is $E(B)$, where $B$ is a random draw from the distribution of $b$. The more heterogeneous the distribution of $b$, the more disparate the results of such covariate-informed and covariate-agnostic decisions will be. This is the basis of our proposed metric.

## Concentration of benefit index ($C_b$)

We start by focusing on the difference between the covariate-informed and covariate-agnostic treatment rules:

$$\Delta_b = E\{max(B_1, B_2)\} - E(B) = \frac{1}{2}.E(|B_1 - B_2|).$$

$\Delta_b$ is a threshold-free quantity that can be estimated for any outcome and arbitrary set of covariates. Its possible values range between 0 and $+\infty$. It has a value of 0 when $var(B) = 0$, indicating that there is no gain in individualizing treatment decisions. Generally, the more dispersed the distribution of $b$, the larger the value of $\Delta_b$.

While $\Delta_b$ is a threshold-free metric, it is still context-specific as it is in the same unit as the benefit of treatment. One approach towards developing a dimensionless metric is to focus on $\Delta_b/E(B)$. This quantity is equal the Gini index for benefit (for a random variable $X$, $Gini_X = \frac{E(|X_1 - X_2|)}{2.E(X_3)}$ where $X_1$, $X_2$, and $X_3$ are independent random draws from the distribution[15]). While the Gini index for strictly non-negative quantities such as risk is well known and is well behaved (e.g., always bounded in [0,1])[16], this is not necessarily the case with treatment benefit. Our main concern is that this quantity can grow without bounds if $E(B)$ is close to zero, resulting in large values that are difficult to interpret.

Instead, our proposed index is a dimensionless metric obtained through dividing $\Delta_b$ by $E\{max(B_1, B_2)\}$, the larger quantity of the two terms comprising $\Delta_b$. We call the resulting quantity the 'Concentration of Benefit' index ($C_b$):

$$C_b = 1 - \frac{E(B)}{E\{max(B_1, B_2)\}} = \frac{Gini_b}{1 + Gini_b}.$$

$C_b$ ranges between (0,1) and can be expressed in percentages: a $C_b$ of p% means that the covariate-agnostic treatment rule is (100-p)% as efficient as the covariate-informed one in the two-subject experiment explained above. When $var(B) = 0$, there is no extra benefit in individualized versus population-based treatment, and $C_b = 0$. If the expected treatment benefit for every individual is non-negative, given that $E\{max(B_1, B_2)\} \leq 2.E(B)$, $C_b$ will be bounded in [0, 0.5]. If there is 'qualitative' interaction such that the expected treatment benefit is positive in some subjects and negative in others, $C_b$ can be more than 0.5. At extreme, $C_b = 1$, indicating that there is no benefit in population-based treatment, but individualized treatment can be beneficial. For example, if the population consists of equal proportion of males and females, and the expected treatment benefit is +a in females and –a in females, $E(B) = 0$ and $C_b = 1$ (and $Gini_b = +\infty$).

## Relationship with treatment threshold

While the concept of giving treatment to one in a pair might seem abstract, it has a firm relationship with treatment rules based on treatment thresholds, such as those explored by VanderWeele et al(5). Imagine we would like to compare the performance of covariate-informed and covariate-agnostic treatment rules when a fraction $p$ of the population would receive the treatment. For the covariate-informed scenario, the most efficient strategy is to provide the treatment to the proportion $p$ of the population with the highest values of $b$. Let $F^{-1}$ be the quantile function of $b$ in the population; this strategy entails setting a treatment threshold at $F^{-1}(1 - p)$ on $b$, such that only subjects with higher expected treatment benefit shall receive the treatment. The population benefit of such a strategy will be

$$benefit(p) = p.E\big(B\big|B > F^{-1}(1-p)\big) = \int_{1-p}^{1} F^{-1}(q).dq.$$

Because we might not know the treatment threshold, we can assume that it can result in any proportion of the population, with equal likelihood, to be treated ($p \sim uniform(0,1)$). As such, the expected (integrated) treatment benefit across the entire range of thresholds is

$$E\{benefit(p)\} = \int_{0}^{1}\left[\int_{1-p}^{1} F^{-1}(q).dq\right].dp = \frac{1}{2}.E\{max(B_1, B_2)\}.$$

On the other hand, in the absence of any information on treatment benefit, the covariate-agnostic rule can do no better (or worse) than randomly assigning a proportion $p$ to treatment, with population benefit of $benefit(p) = p.E(B)$. The integrated benefit of treatment for the covariate-agnostic rule across all $p$ is therefore $\frac{1}{2}E(B)$. Given these equalities, $1 - C_b$ is the average efficiency of covariate-agnostic versus covariate-informed treatment rule across the entire range of treatment thresholds.

## Estimation

For a trial of size $n$, we define $\hat{\mathbf{b}} = \{\hat{b}_1 \geq \hat{b}_2 \geq \hat{b}_3, \ldots, \hat{b}_n\}$ as the ordered, from large to small, vector of estimated treatment benefits. Here we propose two classes of estimators for $C_b$.

### *Parametric (model-based) estimator*

One approach is to directly work with the model-based estimates of expected treatment benefits for each subject. Indeed, $\hat{E}(B) = \frac{1}{n}.\sum_{i=1}^{n}\hat{b}_i$. For $E\{max(B_1, B_2)\}$, given that in the ordered vector $\hat{\mathbf{b}}$, $\forall i \leq j$, $\max(\hat{b}_i, \hat{b}_j) = \hat{b}_i$, instead of averaging across all pairs in the sample, we can proceed as

$$\hat{E}\{max(B_1, B_2)\} = 2\{\frac{1}{n^2} \cdot \sum_{i=1}^{n} \hat{S}(i) - \frac{1}{n} \cdot \hat{E}(B)\},$$

where $\hat{S}(k) = \sum_{i=1}^{k} \hat{b}_i$ is the partial sum of $\hat{\mathbf{b}}$ up to and including its k$^{th}$ element.

This purely model-based approach for estimating $C_b$ cannot be used for evaluating the out-of-sample performance of the model. Imagine the prediction model for expected treatment benefit in the original sample is to be evaluated in a new clinical trial data. The prediction model can be used to estimate $\hat{b}_i$ for each new subject based on their observed covariates, and therefore allows estimating $C_b$; but such an approach is independent of the observed outcomes in the new sample. This limits evaluating the external validity of estimates and investigation of the level of optimism of the estimand.

### *Semi-parametric estimator*

Here we propose an alternative semi-parametric estimation method that is influenced by the actual outcomes in the new sample. Our derivations closely follows that of Zhao et al(17). We note that a consistent estimator for $E(B)$ in the new sample with random treatment assignment is the difference in the mean of observed outcomes between the two groups. For example, for a binary outcome in the absence of censoring,

$$\hat{E}(B) = \frac{\sum_{i=1}^{n} I(A_i = 0).Y_i}{\sum_{i=1}^{n} I(A_i = 0)} - \frac{\sum_{i=1}^{n} I(A_i = 1).Y_i}{\sum_{i=1}^{n} I(A_i = 1)}.$$

Similarly, for $E\{max(B_1, B_2)\}$, we can replace the estimate of the partial sum of expected treatment benefits up to the k$^{th}$ subject in the equation for parametric estimator with the observed difference in the outcomes between the two arms among the first k subjects with the highest predicted treatment benefit:

$$\hat{S}(k) = k \cdot \frac{\sum_{i=1}^{k} I(A_i = 0).Y_i}{\sum_{i=1}^{k} I(A_i = 0)} - k \cdot \frac{\sum_{i=1}^{k} I(A_i = 1).Y_i}{\sum_{i=1}^{k} I(A_i = 1)}.$$

We note that for the first few values of k, the estimator might result in undefined values. For example, for k=1 (the first subject), either $A_1 = 0$ or $A_1 = 1$, and the other term is 0/0. In the implementation of this estimator, we have carried backward the first defined value within each treatment group. The practical impact of different approaches to overcome this issue should only be of concern in small samples.

We consider this method semi-parametric as it is still based on a (parametric) model for $E(Y|\mathbf{X}, A)$, but the model acts as a ranking mechanism for observations, and the estimate of cumulative benefit is obtained non-parametrically from the ranked vector of observed outcomes.

**Quantifying uncertainty**

Like any summary statistic estimated form a finite sample, $C_b$ has a sampling distribution. The interpretation of $C_b$ as the relative performance of individualized versus population-based treatment across all treatment thresholds provides a decision-theoretic perspective which makes inference less relevant(18). However, the end-user might still be interested in examining the range of values that are compatible with the data. Closed-form equations might exist for variance functions assuming certain parametric distribution for $b$. However, it is important to incorporate uncertainty in the model structure. For constructing confidence intervals for the general case, we suggest a bootstrapping approach in which all the calculations (including the estimation of the shrinkage penalty term) are repeated within bootstrapped samples.

**Application: a clinical trial of preventive antibiotic therapy for Chronic Obstructive Pulmonary Disease (COPD) exacerbations**

COPD is a chronic airways disease characterized by the loss of lung function and chronic symptoms such as cough and shortness of breath. Exacerbations of COPD are periods of intensified disease activity and are associated with high risk of morbidity and mortality and increased use of healthcare resources(19). The MACRO study was a randomized trial of daily azithromycin, a broad-spectrum antibiotic, for the prevention of acute exacerbations of COPD(20). In this study, patients were randomized to either azithromycin (n=570) or placebo (n=572) and were followed for up to one year. The rate of exacerbations was 1.48 per patient-year in the azithromycin group, as compared with 1.83 in the placebo group (p=0.01). A proportional hazards model for the time to the first exacerbation demonstrated a 27% reduction in the rate of exacerbations in the azithromycin relative to the placebo group(20). However, using azithromycin will incur costs, and there are concerns around azithromycin therapy including the risk of adverse cardiovascular events(21). Targeting the subset of the COPD patients who will likely gain the most benefit from this therapy can be a more efficient strategy than treating all patients.

In this context, a natural definition of benefit is the number of exacerbations avoided due to treatment over one unit of time, which we define to be one year. For the purpose of this example we focus on moderate (requiring outpatient care) and severe (requiring inpatient care) exacerbations. Mild exacerbations for which the patient will not use healthcare resources were not deemed relevant for preventive therapy. Because reduction in the absolute number of exacerbations is the outcome of interest, instead of the non-parametric proportional hazards model in the main analysis, we switch to a parametric count model of the outcome that considers all events during follow-up.

As the subgroup-defining variables we focus on the following six covariates: sex, age at baseline, whether the patient was hospitalized due to a COPD exacerbation in the 12-month period before enrollment, whether the patient received systemic corticosteroids due to a COPD exacerbation in the 12-month period before enrollment, forced expiratory volume at one second ($FEV_1$ – a measure of lung function) at baseline, as well as the baseline St. George Respiratory Questionnaire (SGRQ) score (a measure of functional capacity, with higher scores indicating lower capacity). These variables were selected *a priori* by an expert clinician as potential predictors of exacerbation rate or modifiers of treatment effect, which can conceivably be used to formulate a prediction score for benefit of preventive therapy. There were 1,108 patients with non-missing covariates, comprising the sample for this analysis. These patients were followed for an average of 0.94 years and contributed 544 events. As in the original analysis, we consider loss-to-follow-up to have occurred at random. A previous analysis demonstrated that the pattern of exacerbation occurrence was consistent with a constant hazard over time(22). Fewer than 4% of individuals died and a previous analysis showed little impact from the competing risk of death on the statistical inference on exacerbation rate(22). As the treatment is randomized, there is no structural confounding. The possibility of chance confounding (noticeable imbalance of covariates due to randomization(23)) was ruled out in an exploratory analysis (the standardized mean difference between treatment groups was less than 5% for all covariates). As such, we fitted a generalized linear model with negative binomial distribution and logarithmic link function to model the expected exacerbation rate as a function of treatment and covariates. We start by a 'full' model that includes both independent effects for $A$ and $\mathbf{X}$, and their interaction for each of the $m$ covariates:

$$\hat{E}(Y_i | A_i, \mathbf{X}_i, T_i) = \exp\left(\beta_0 + \beta_a . A_i + \sum_{j=1}^{m} \beta_j . X_{i,j} + \sum_{j=1}^{m} \beta_{a.j} . A_i . X_{i,j} + \ln(T_i)\right).$$

Here, $\mathbf{X} = \{X_{i,j} : i = 1,2, \dots, n; j = 1,2, \dots, m\}$ is vector of $m = 6$ covariates for the i[th] subject, $Y_i$ is the number of exacerbations during follow-up, $A_i$ indicates treatment (1:daily azithromycin, 0:placebo), and $T_i$ is the follow-up time in years, whose logarithm enters the model as the offset variable.

To estimate the parameters of this model, we chose the ridge regression in the main analysis, and unconstrained ML in a secondary analysis. Ridge applies an $\ell_2$ norm penalty of $\lambda . \sum \beta^2$ to the likelihood function (when all independent variables are standardized) to prevent optimistic predictions(24). We used 10-fold cross-validation to find the optimal value of $\lambda$ that minimized the mean-squared error of off-sample predictions. All calculations were performed in R(25); we used the implementation of ridge regression provided in the *glmnet* package(26). We note that other popular regularization techniques, such as the least absolute shrinkage and selection operator (Lasso) can also be used(27). Lasso applies an $\ell_1$ norm penalty with the property that some coefficients can be shrunk to exactly zero, producing parsimonious prediction models; however, this feature of Lasso is less relevant to the present context, where the emphasis is on global predictive performance of covariates rather than publicizing any treatment rule.

**Table 1** provides the estimates of regression coefficients for both ridge and ML models, after all variables were standardized to have zero mean and unit variance. Using the ridge regression resulted in the shrinkage of the majority of variables towards zero.

**Table 1:** Regression coefficients for the ridge and maximum likelihood estimates

| Variables | Ridge regression estimates | Unconstrained ML estimates |
|---|---|---|
| **Intercept** | -1.367 | -1.959 |
| **Main effects** | | |
| **Treatment (1:azithromycin, 0:placebo)** | -0.116 | 0.693 |
| **Sex (female v. male)** | -0.155 | -0.241 |
| **Age (in years)** | -0.008 | -0.004 |
| **History of previous hospitalization due to COPD (binary)** | 0.744 | 0.976 |
| **History of previous use of systemic corticosteroids (binary)** | 0.420 | 0.479 |
| **Baseline forced expiratory volume at one second (FEV$_1$)** | -0.108 | -0.176 |
| **SGRQ score*** | 1.395 | 1.792 |
| **Treatment by covariate interaction effects** | | |
| **Sex** | 0.017 | 0.109 |
| **Age** | -0.003 | -0.014 |
| **History of previous hospitalization due to COPD** | 0.172 | 0.045 |
| **History of previous use of systemic corticosteroids** | 0.005 | 0.121 |
| **Baseline FEV$_1$** | -0.063 | 0.006 |
| **SGRQ score*** | -0.118 | -0.555 |

*All variables are scaled to have zero mean and unit variance

***ML: ;*** maximum likelihood; ***COPD:*** Chronic Obstructive Pulmonary Disease; ***FEV$_1$:*** forced expiratory volume at one second; ***SGRQ:*** St. George's Respiratory Questionnaire

**Figure 1** demonstrates the histogram of estimated expected treatment benefits ($\hat{b}$) for each subject based on the reference model. On average, one year of treatment with azithromycin prevented 0.153 exacerbations. However, there was a marked variability in this number across the sample, with the interquartile range being $0.108 - 0.187$.

**Figure 1:** histogram of expected treatment benefit as a function of baseline covariates

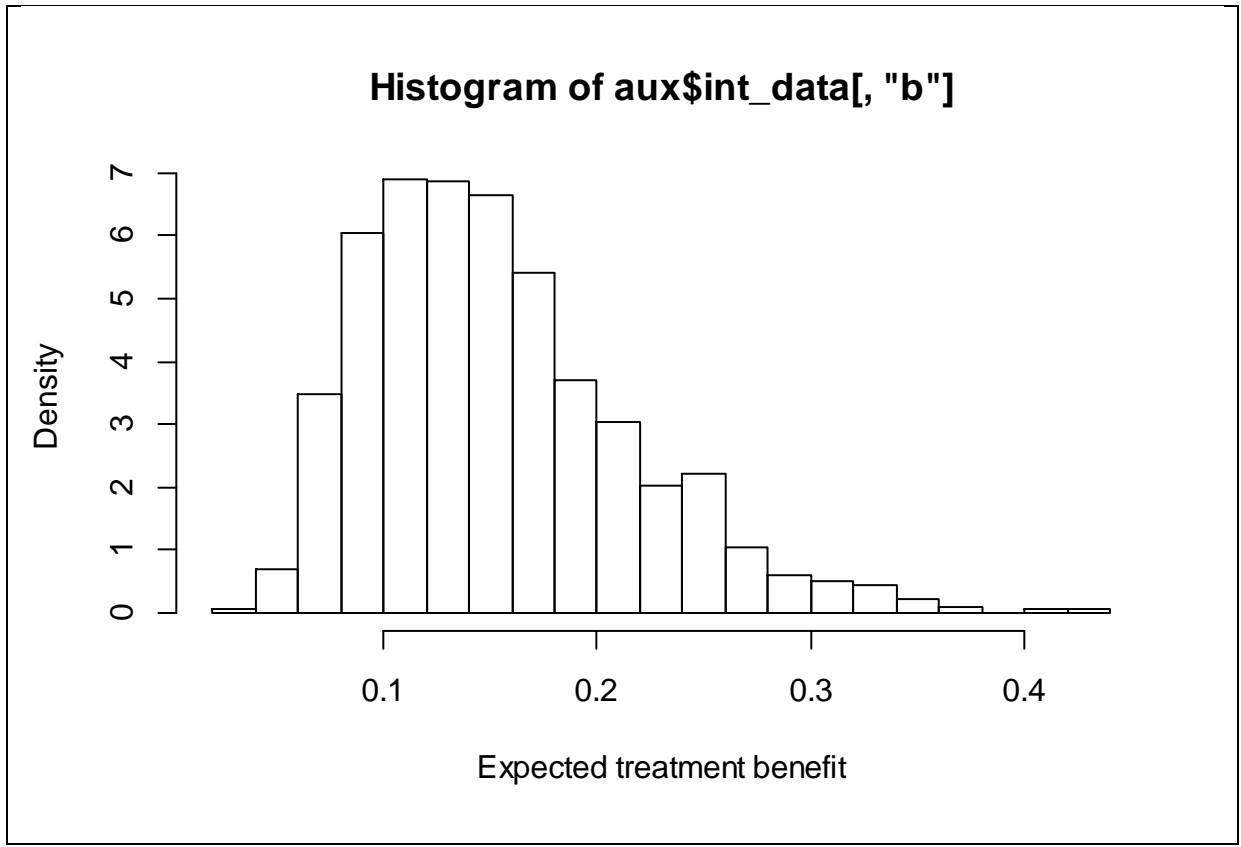

For both ridge and ML models, we evaluate the parametric and semi-parametric estimations of $C_b$. Inference was based on 1,000 bootstraps, with parameter estimation and shrinkage being repeated within each bootstrapped sample.

Note that the presence of variable follow-up times requires the incorporation of time in the semi-parametric estimators. For $\hat{E}(B)$, the estimate is the difference in the annual exacerbation rate between the two groups. Similarly, for $\hat{E}\{max(B_1, B_2)\}$, the partial sum component would be

$$\hat{S}(k) = k . \frac{\sum_{i=1}^{k} I(A_i = 0) . Y_i}{\sum_{i=1}^{k} I(A_i = 0) . T_i} - k . \frac{\sum_{i=1}^{k} I(A_i = 1) . Y_i}{\sum_{i=1}^{k} I(A_i = 1) . T_i} .$$

The performance of the parametric versus semi-parametric estimators of $S(.)$ fcan be evaluated visually, as in ***Figure 2***.

**Figure 2:** Running partial sum of model-estimated $b$ (continuous line) and its semi-parametric estimator (jagged line) for the ridge regression.

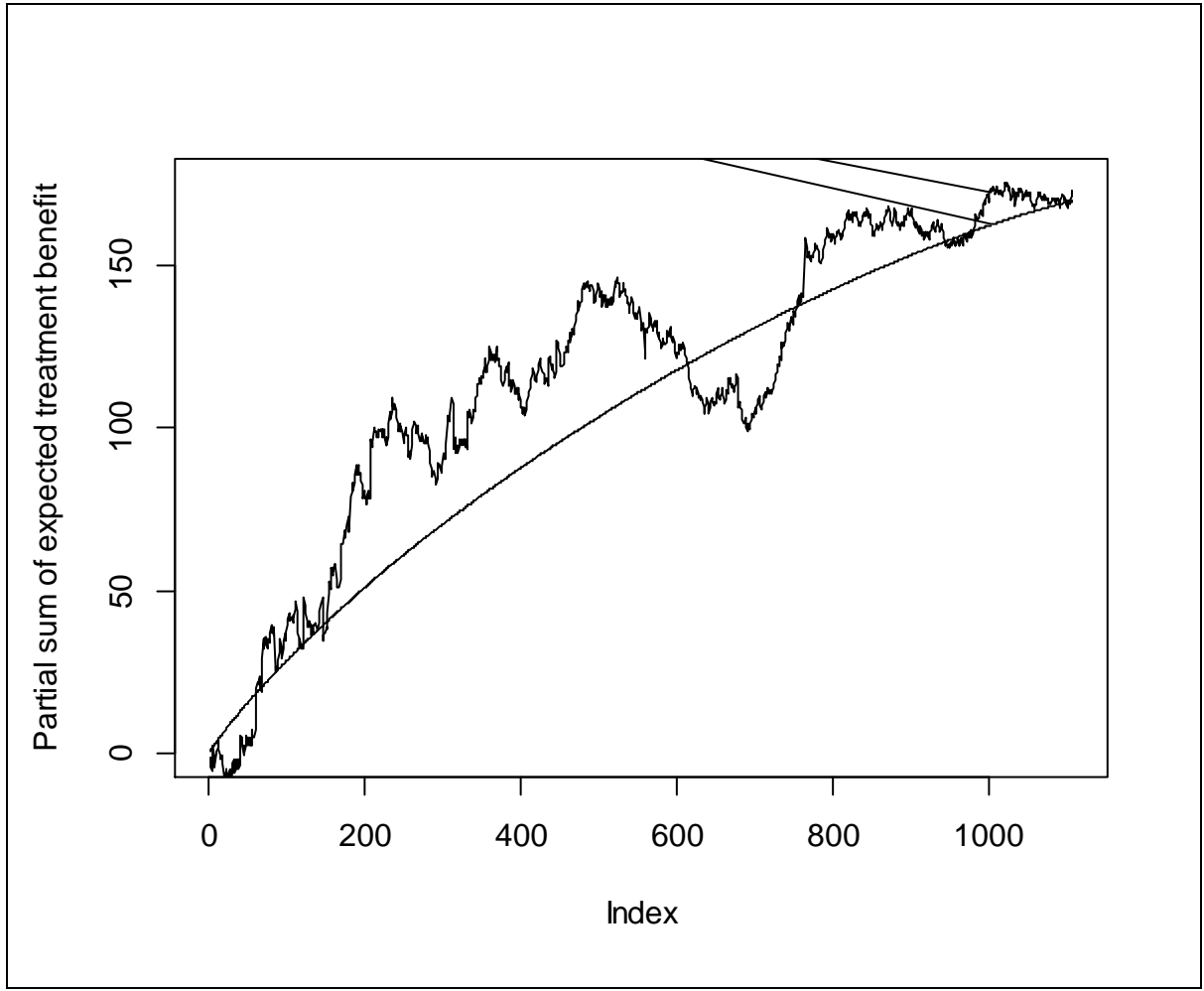

The parametric estimate of $C_b$ based on the ridge regression model was 0.171 (95%CI $0.151 - 0.587$), indicating that the population-level treatment is 82.9% as efficient as covariate-informed treatment on average across all treatment thresholds. The ML-based model yielded a value of 0.307 (95%CI $0.251 - 0.640$). The semiparametric estimation yielded values of 0.249 (95%CI $0.137 - 0.761$) for the ridge model and 0.275 (95%CI $0.187 - 0.701$) for the ML model.

## A brief simulation study

To evaluate the behavior of the proposed indices, a series of simulations were conducted. The empirical joint distribution of the six covariates from the case study was taken as their population distribution to generate a super-population of 10^7 simulated individuals. We modeled the true treatment effect under three scenarios: 1) strong interaction on the log scale, in which the ML estimates of the regression (Table 1) were used to generate simulated treatment outcomes; 2) weak interaction on the log scale, in which the ridge estimates (Table 1) were used; and 3) a null model, in which treatment effect was modeled not to be affected by any of the covariates. The population values of $C_b$ were as follows: 0.334 for the strong interaction scenario, 0.183 for the weak interaction scenario, and 0 for the null scenario.

For each scenario, we generated simulated clinical trials of small (n=400), medial (n=1,000), and large (n=5,000) sample sizes. Treatment was assigned to each subject at random with a probability of 0.5. Within each simulation loop, we fitted the penalized and unconstrained ML models (with the same structure as in the case study). Optimism adjustment was based on 200 bootstraps. A new model was fitted (including shrinkage estimation) in the bootstrapped sample (within-sample estimate), and was used to calculate the indices in the original sample (out-of-sample estimate). The difference between within-sample and out-of-sample estimates was averaged over bootstrap iterations. Results, based on 1,000 simulations, are provided in **_Table 3_**.

**_Table 3:_** Results of the simulations

| N | Parametric | | | | | | Semi-parametric | | | | | | Semi-parametric, optimism adjusted | | | | | |
|---|---|---|---|---|---|---|---|---|---|---|---|---|---|---|---|---|---|---|
| | Ridge | | | ML | | | Ridge | | | ML | | | Ridge | | | ML | | |
| | _Bias_ | _SD_ | _RMSE_ | _Bias_ | _SD_ | _RMSE_ | _Bias_ | _SD_ | _RMSE_ | _Bias_ | _SD_ | _RMSE_ | _Bias_ | _SD_ | _RMSE_ | _Bias_ | _SD_ | _RMSE_ |
| **Strong interaction on the log scale** (see Table 1 for regression coefficients for treatment effect; $C_b = 0.334$) | | | | | | | | | | | | | | | | | | |
| **400** | -0.018 | 0.130 | 0.131 | 0.120 | 0.132 | 0.178 | 0.086 | 0.168 | 0.189 | 0.122 | 0.161 | 0.202 | -0.034 | 0.204 | 0.207 | 0.000 | 0.207 | 0.207 |
| **1,000** | -0.047 | 0.065 | 0.081 | 0.051 | 0.072 | 0.088 | 0.032 | 0.091 | 0.096 | 0.051 | 0.091 | 0.104 | -0.051 | 0.103 | 0.115 | -0.035 | 0.106 | 0.112 |
| **5,000** | -0.063 | 0.024 | 0.068 | 0.012 | 0.031 | 0.034 | -0.001 | 0.039 | 0.039 | 0.014 | 0.038 | 0.041 | -0.019 | 0.04 | 0.045 | -0.008 | 0.04 | 0.041 |
| **Weak interaction on the log scale** | | | | | | | | | | | | | | | | | | |
| (see Table 1 for regression coefficients for treatment effect; $c_b = 0.183$) | | | | | | | | | | | | | | | | | | |
| **400** | 0.095 | 0.146 | 0.174 | 0.249 | 0.143 | 0.287 | 0.196 | 0.183 | 0.268 | 0.246 | 0.171 | 0.299 | 0.053 | 0.215 | 0.221 | 0.086 | 0.222 | 0.238 |
| **1,000** | 0.046 | 0.080 | 0.092 | 0.142 | 0.09 | 0.167 | 0.109 | 0.114 | 0.157 | 0.139 | 0.108 | 0.176 | -0.011 | 0.124 | 0.124 | 0.001 | 0.127 | 0.126 |
| **5,000** | 0.006 | 0.029 | 0.03 | 0.039 | 0.041 | 0.056 | 0.03 | 0.051 | 0.059 | 0.039 | 0.049 | 0.062 | -0.019 | 0.061 | 0.064 | -0.022 | 0.062 | 0.066 |
| **Null model** | | | | | | | | | | | | | | | | | | |
| (No heterogeneity of treatment effect; $c_b = 0$) | | | | | | | | | | | | | | | | | | |
| **400** | 0.102 | 0.067 | 0.122 | 0.396 | 0.186 | 0.296 | 0.230 | 0.186 | 0.296 | 0.394 | 0.136 | 0.417 | 0.133 | 0.187 | 0.229 | 0.221 | 0.186 | 0.289 |
| **1,000** | 0.085 | 0.036 | 0.092 | 0.284 | 0.092 | 0.299 | 0.146 | 0.120 | 0.189 | 0.282 | 0.094 | 0.297 | 0.068 | 0.098 | 0.119 | 0.133 | 0.105 | 0.169 |
| **5,000** | 0.073 | 0.02 | 0.076 | 0.149 | 0.041 | 0.155 | 0.072 | 0.060 | 0.094 | 0.148 | 0.043 | 0.154 | 0.033 | 0.045 | 0.056 | 0.064 | 0.050 | 0.081 |

*All simulations are run for 1,000 times.*

*ML: maximum likelihood; SD: standard deviation; RMSE: Root mean square error*

Overall, $\hat{C}_b$ generally behaved as expected for a consistent estimator, shrinking in bias, SD, and RMSE with larger sample sizes. The parametric ML estimator was upwardly biased, which could be severe in small sample sizes. Penalized regression moderately underestimated $C_b$ in the strong interaction scenario and overestimated it in the weak interaction scenario. Generally, penalized regression had lower dispersion (SD) than its ML counterpart, while semi-parametric estimators had generally higher dispersion than parametric ones. In the presence of covariate-by-treatment interaction, optimism adjustment generally removed the bias but increased the SD. There was no discernible difference between the ridge and ML estimators in both strong and weak interaction scenarios after optimism correction. Given that $C_b$ has its minimum population value under the null scenario (when outcome is not a function of covariates), it is expected that estimates would be upwardly biased. For the ML estimators, this bias persisted even with large sample sizes. Overall, judging by the root mean square error, the parametric model with shrinkage estimation prevailed in seven of the nine scenarios (with the two exceptions being parametric ML estimator without shrinkage for the strong interaction scenario, and optimism-adjusted semi-parametric ridge in the null scenario, outperforming it, both when N=5,000).

**Discussion**

We presented a novel metric that summarizes the capacity of subgroup-defining variables towards formulating efficient treatment rules. The metric quantifies such capacity based on quantifying improvement in the expected efficiency of treatment when a covariate-informed treatment rule is used in place of a covariate-agnostic rule to decide which of two randomly selected subjects should receive treatment. We showed that the resulting index, $C_b$, is closely related to the Gini index for the distribution of expected treatment benefit in the population. Importantly, we showed that $C_b$ can also be interpreted as evaluating the benefit of covariate-informed versus covariate-naive treatment decisions, integrated over the range of treatment thresholds. Through a case study, we demonstrated how such calculations can be performed in the context of a trial in which the outcome was a count event. This approach can easily be used with a variety of inference techniques that are used in the analysis of clinical trials, such as logistic regression or proportional hazard and accelerated failure time models. For example, empirical estimates of baseline hazard and hazard ratios from a proportional hazards model can be used to estimate cumulative incidence of the outcome and the absolute rate reduction up to a time horizon of interest, which will provide sufficient data to calculate the related metrics.

Our proposed approach is in line with recent recommendations for direct evaluation of treatment benefit as a function of covariates(4). A major advantage of such treatment benefit modeling is that covariates, and their interaction with treatment, can be considered jointly. In this case, the $C_b$ can be seen as a global metric of the combined capacity of covariates towards enabling personalized treatment decisions. A similar approach has recently been proposed by van Klaveren et al(28). Their proposed 'c-for-benefit' represents the probability that from two randomly chosen pairs of individuals with unequal observed benefit, the pair with higher observed benefit also has a higher predicted benefit. In contrast, 1-$C_b$ is the relative efficiency of providing treatment to one of two randomly selected subjects at random, as opposed to providing treatment to one with higher expected benefit. Comparison between the behaviors of these metrics remains to be evaluated in future studies.

The interpretation of $C_b$ as the benefit of covariate-informed treatment integrated over treatment thresholds allows one to devise modified versions of this index that pertain to a specific range on the treatment threshold. In certain instances (e.g., a positive trial for a treatment with low rate of adverse events), we expect that the treatment will be provided to the majority of the population, and covariates are to be used to find the minority for which treatment should be opted out. In other instances (e.g., a negative trial), the investigator might be in search for the minority in the population who should opt in for treatment. Alternatively, one can propose a minimally acceptable treatment benefit (or a maximally acceptable number-needed-to-treat) given the risk profile of treatment, which will define a boundary on the treatment threshold. The 'partial' $C_b$ calculated in this way is similar in concept to the partial AUC of the ROC curve(29).

The finite sample properties of the proposed index, and its sensitivity to model specification and various regularization techniques should be evaluated in future studies. Both the case study and the simulations showed that $C_b$ is sensitive to the choice of parametric versus semiparametric estimation, whether shrinkage estimation is employed, and whether estimates are corrected for optimism. Recent simulation studies also point towards the sensitivity of estimates of treatment effect heterogeneity to such specifications even in large sample sizes(30). In our simulations, parametric estimation with shrinkage resulted in the estimates with the lowest average root mean square error, but this needs to be further evaluated. In general, model specification and the choice of estimation method will require careful examination of model fit. When reporting clinical trial results, full pre-specification of the modeling approach can add to the credibility of estimates and consistency across studies.

Other research on this topic can focus on how $C_b$ can be estimated from observational studies. In our case study, randomization had protected against systematic confounding, and the relatively large sample size protected against chance (non-structural) imbalance of confounders. However, if this condition cannot be assured, the potential outcome models must include the confounding variables. For example, if $\mathbf{X}$, the set of covariates of interest, is a subset of a larger set $\mathbf{X}^*$ that is required to block the confounding (backdoor) paths, then the expected treatment benefit should be calculated as $b = E_{\mathbf{X}^*}\{E(Y|\mathbf{X}, A = 0) - E(Y|\mathbf{X}, A = 1)\}$. Evaluating this nested expectation remains an important topic to investigate. Another research topic is applying this concept to multi-arm trials. One extension of the two-subject treatment assignment task underlying $C_b$ to where there are $m$ treatment options can be based on comparing the outcome of covariate-informed and covariate-agnostic rules when assigning each of $m$ treatments to each of $m$ randomly selected subjects.

## Conclusion

Subgroup analysis in clinical trials is an early opportunity for exploring the potential of using patient characteristics to move from population-level treatment decisions to more efficient individualized treatment rules. While decision theory provides a consistent framework for this problem, its application requires difficult-to-establish treatment thresholds. Our contribution in this work was the introduction of global measures of the capacity of covariates in designing efficient treatment rules. To us, this is the 'sweet spot' for subgroup analysis when clinical trial results are being reported. Several important issues around model selection, variable election, and extension to observational studies and multi-arm trials remain to be explored.

## Acknowledgement

The authors would like to thank Drs. Abdollah Safari and Mahyar Etminan, The University of British Columbia, for their insightful comments on earlier drafts.

# Data Sharing Statement

The data that support the findings of this study are available upon request from the COPD Clinical Research Network (http://copdcrn.org/). Restrictions apply to the availability of these data, which were used under license for this study.